\documentclass[twocolumn,showpacs,aps,prl,superscriptaddress]{revtex4}

\usepackage{graphicx}
\usepackage{dcolumn}
\usepackage{epsfig}
\usepackage{amsmath}
\usepackage{subfigure}

\newcommand{\BaBarYear}    {07}
\newcommand{\BaBarNumber}  {0xx}

\newcommand{\SLACPubNumber} {xxxxx}

\input pubboard/babarsym
\RequirePackage{xspace}

\newcommand{\pvec}{{\bf p}}

\newcommand{\timesix}{\ensuremath{\times10^{6}}}

\newcommand{\DE}{\ensuremath{\Delta E}}

\newcommand{\xf}{\ensuremath{{\cal F}}}

\newcommand\etal{{\it et al.}}
\newcommand{\half}{\ensuremath{{\frac{1}{2}}}}

\newcommand{\bfig}{\begin{figure}[htbpc!]}
\newcommand{\efig}{\end{figure}}
\newcommand\bef{\begin{figure}}
\newcommand\edf{\end{figure}}

\newcommand\beq{\begin{equation}}
\newcommand\eeq{\end{equation}}
\newcommand\bear{\begin{array}}
\newcommand\enar{\end{array}}
\newcommand\beqa{\begin{eqnarray}}
\newcommand\eeqa{\end{eqnarray}}
\newcommand\ben{\begin{enumerate}}
\newcommand\een{\end{enumerate}}

\newcommand{\UfourS}{\ensuremath{\Upsilon(4S)}}

\begin{document}

\preprint{\babar-PUB-\BaBarYear/\BaBarNumber} 
\preprint{SLAC-PUB-\SLACPubNumber} 

\begin{flushleft}
\babar-PUB-07/035\\
SLAC-PUB-12705\\
\end{flushleft}

\par\vskip .2cm

\title{
 \large \bf\boldmath Search for the rare charmless hadronic decay $B^{+} \to a_{0}^{+}\pi^{0}$
}
%
\author{B.~Aubert}
\author{M.~Bona}
\author{D.~Boutigny}
\author{Y.~Karyotakis}
\author{J.~P.~Lees}
\author{V.~Poireau}
\author{X.~Prudent}
\author{V.~Tisserand}
\author{A.~Zghiche}
\affiliation{Laboratoire de Physique des Particules, IN2P3/CNRS et Universit\'e de Savoie, F-74941 Annecy-Le-Vieux, France }
\author{J.~Garra~Tico}
\author{E.~Grauges}
\affiliation{Universitat de Barcelona, Facultat de Fisica, Departament ECM, E-08028 Barcelona, Spain }
% ...
\collaboration{The \babar\ Collaboration}
\noaffiliation

\date{\today}

\begin{abstract}
We present a search for $B$ decays to a charged scalar meson
$a_{0}^{+}$ and a \piz where the $a_{0}^{+}$ decays to an
$\eta$ meson and a \pip. The analysis was performed on a
data sample consisting of 383\timesix\ \BB\ pairs collected with the
\babar\ detector at the PEP-II asymmetric-energy $B$ Factory at SLAC.
We find no significant signal and set an upper limit on the product
branching fraction \\${\cal{B}}(B^{+}\to
a_{0}^{+}\piz)\times{\cal{B}}(a_{0}^{+}\to\eta\pip)$ of
1.4$\times$10$^{-6}$ at the 90\% confidence level.

\end{abstract}


\maketitle

\begin{figure}[!ht] 
\begin{center}
\begin{picture}(100,220)(73,-95)
\put(0,15){\includegraphics[width=0.48\linewidth]{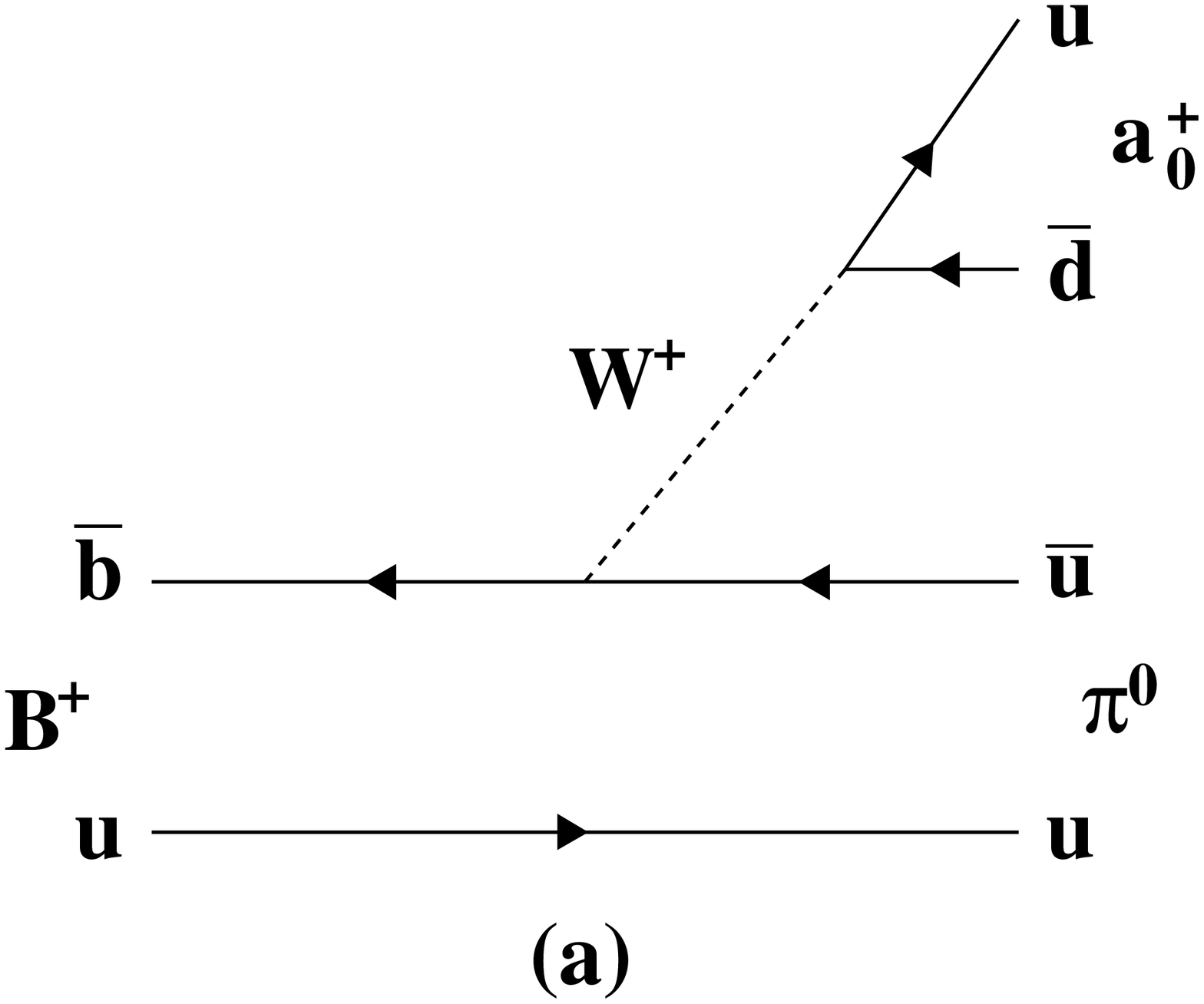}}
\put(125,15){\includegraphics[width=0.48\linewidth]{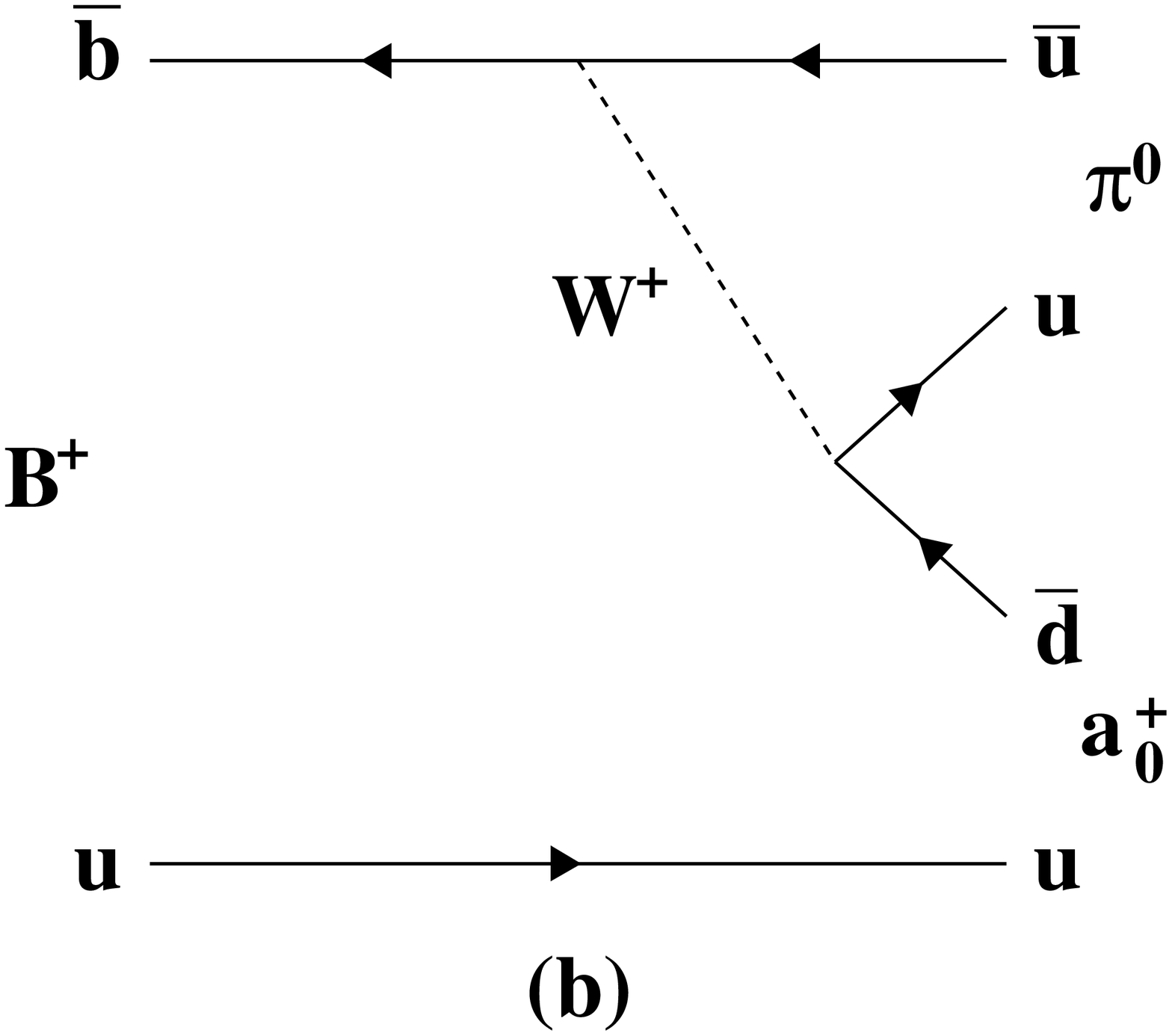}}
\put(0,-95){\includegraphics[width=0.48\linewidth]{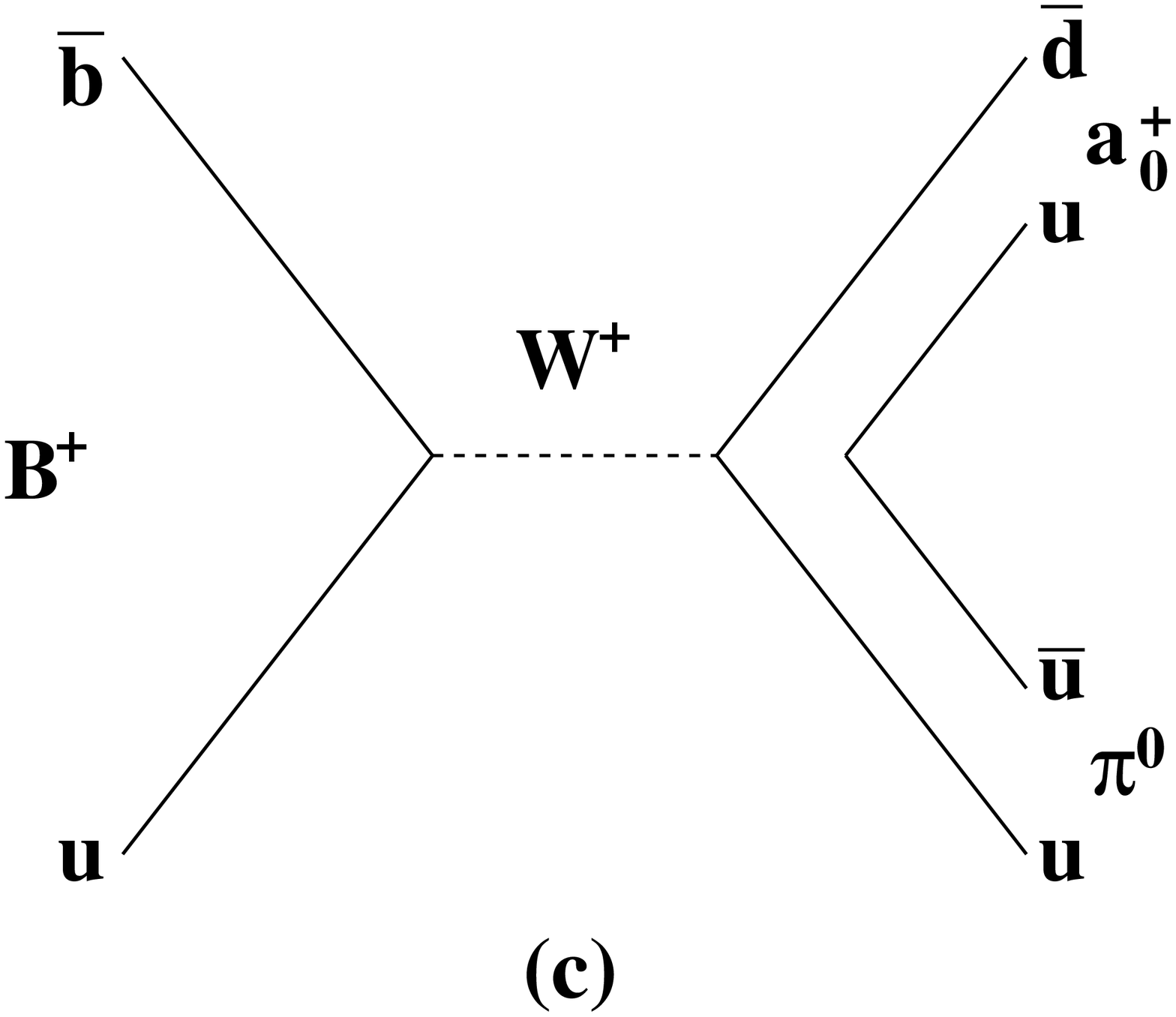}}
\put(125,-95){\includegraphics[width=0.48\linewidth]{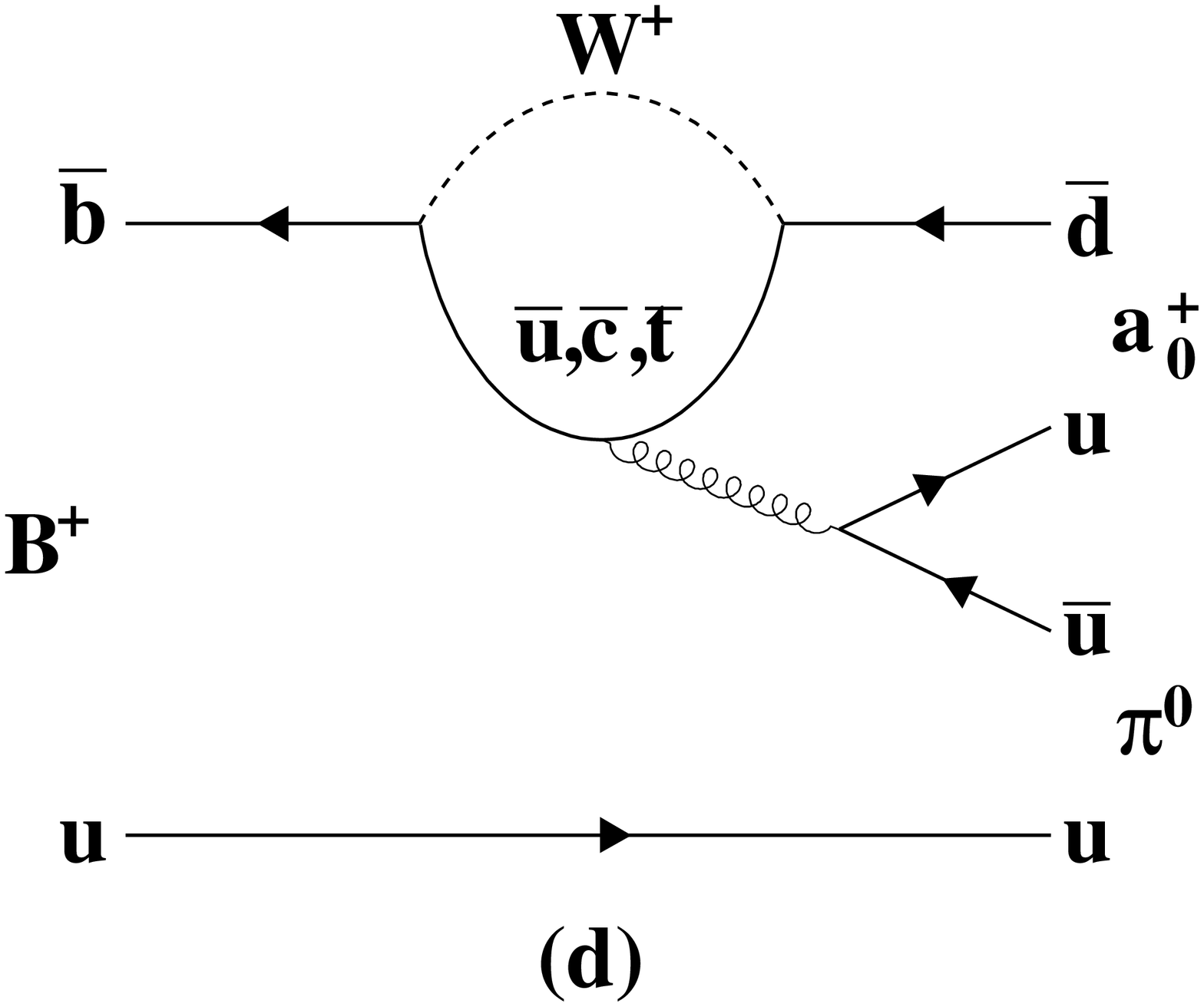}}
\end{picture}
\caption{The Feynman diagrams contributing to the process $B^{+} \to
a_{0}^{+}\piz$ in the two-quark model. (a) is the external
(color-allowed) tree, (b) the internal (color-suppressed) tree, (c)
the annihilation process and (d) the gluonic penguin process.}
\label{feyn:a0} 
\end{center} 
\end{figure}

The structure of scalar mesons is a subject of some
debate~\cite{PDG2006_scalar,baru}. Proposed models include two-quark
or four-quark states with potential contributions from glueballs or a
molecular admixture of $K\Kbar$ meson pairs. Measurement of the
branching fraction for the mode $B^{+} \to a_{0}^{+}\pi^{0}$~\cite{CC}
is expected to provide an effective test of the two- and four-quark
models~\cite{Delepine}. The Feynman diagrams for the decay in the
two-quark case are shown in Figure~\ref{feyn:a0}. Those for the
four-quark case are similar except for an \ssbar pair produced from
the vacuum internal to the $a_{0}^{+}$ meson. The color-allowed
electroweak tree diagram shown in Figure~\ref{feyn:a0}(a) is
suppressed for all $a_{0}^{+}$ models since the $W^{+}$ is constrained
to decay to states of even $G$-parity (a generalization of $C$
symmetry to cover particle multiplets) within the Standard Model,
whereas the $a_{0}^{+}$ has odd $G$-parity~\cite{vasia}. This diagram
is also suppressed due to vector current conservation
considerations. Therefore, the color-suppressed tree diagram in
Figure~\ref{feyn:a0}(b) and the helicity-suppressed electroweak
annihilation diagram in Figure~\ref{feyn:a0}(c) become important. The
gluonic penguin process in Figure~\ref{feyn:a0}(d) is highly
suppressed and is therefore not expected to contribute significantly.

The amplitudes for the above diagrams depend on the $a_{0}^{+}$ model
used; in particular the annihilation diagram is heavily suppressed in
a four-quark model. Hence measurement of the branching fraction
provides the potential for model discrimination. In the two-quark
case, the predicted branching fractions go as high as
2$\times$10$^{-7}$~\cite{Delepine,cheng}. However, in the four-quark
case the prediction for the branching fraction is an order of
magnitude lower.

The branching fraction for the result quoted below will be given in
terms of the product ${\cal{B}}(B^{+}\to
a_{0}^{+}\piz)\times{\cal{B}}(a_{0}^{+}\to\eta\pip)$ since the
branching fraction ${\cal{B}}(a_{0}^{+}\to\eta\pi^{+})$ is not well
measured, although it is thought to be approximately
85$\%$~\cite{PDG2006_scalar}.

The analysis presented in this paper is based on 347~fb$^{-1}$
of data collected at the $\Upsilon (4S)$ resonance with the \babar\
detector at the PEP-II asymmetric-energy $e^+e^-$
collider located at the Stanford Linear Accelerator Center. This
corresponds to (383$\pm$4)\timesix\ \BB\ pairs.

The $\babar$ detector has been described in detail
previously~\cite{BABARNIM}. Track parameters of charged particles are
measured by a combination of a 5-layer double-sided silicon vertex
tracker and a 40-layer drift chamber (DCH), both operating in the 1.5~T
magnetic field of a superconducting solenoid. Photons and electrons
are identified using a CsI(Tl) electromagnetic calorimeter.
Further charged particle identification (PID) is provided by
measurements of the average energy loss (\dedx) in the tracking
devices and by an internally-reflecting, ring-imaging \v{C}erenkov
detector (DIRC) covering the central region.

The analysis focuses on $a_{0}^{+}$ mesons produced from the decay
$B^{+} \to a_{0}^{+}\piz$, followed by $a_{0}^{+}\to\eta\pi^{+}$,
where the $\eta$ meson subsequently decays to $\gamma\gamma$ or
$\pip\pim\piz$ final states. The \piz mesons used are reconstructed
via the decay $\piz\to\gamma\gamma$. The selections used for the
analysis are the result of an optimization procedure based on ensemble
Monte Carlo (MC) studies. In these studies, a sample of MC candidates
is produced for given selection criteria by generating randomly from
probability density function (PDF) distributions defined with the
selection applied. By re-fitting to the datasets for each set of
selection criteria it is possible to select the set that yields the
maximum sensitivity to signal. This is done independently for each
decay mode considered. In both cases $a_{0}^{+}$ candidates are
required to satisfy $0.8 < m_{\eta\pi} < 1.2$\gevcc with the $\eta$
candidates satisfying $0.51< m_{\gamma\gamma}<0.57$\gevcc or
$0.540<m_{3\pi}<0.555$\gevcc. The \piz produced from the
$\eta\to\pip\pim\piz$ decay is required to satisfy $0.10 < m_{\piz} <
0.16$\gevcc. The \piz daughter of the $B$ candidate is required to
satisfy $0.115 < m_{\piz} < 0.150$\gevcc. This selection is tighter
than for the \piz produced from the $\eta$ meson since it is of
significantly higher energy and therefore has a better resolution. The
charged track from the $a_{0}^{+}$ candidate decay is required not to
satisfy DIRC- and DCH-based PID criteria consistent with a kaon
hypothesis. This PID selection has been measured to be more than
80$\%$ efficient for tracks with momenta up to 4\gevc with a pion
mis-identification rate lower than 10$\%$ over the same range.

A $B$ meson candidate is characterized kinematically by the energy
difference $\DE \equiv E_B-\half\sqrt{s}$ and energy-substituted mass
$\mes \equiv (\frac{1}{4}s-\pvec_B^2)^\half$, where $s$ is the square
of the centre-of-mass energy of the colliding beams, $(E_B,\pvec_B)$
is the candidate $B$ meson 4-momentum and all values are expressed in
the \UfourS\ frame.  Signal events peak around zero for $\DE$, and at
the $B$ meson mass for $\mes$. The resolutions for $\DE$ and $\mes$
are approximately 30\mev and 3\mevcc, respectively. We require
$|\DE|\le0.35$\gev and $5.20\le\mes\le5.29\ \gevcc$ as an input for
the fit used to extract signal and background parameters (described
below) in order to maximize the available statistics.

The principal source of background in the analysis arises from random
combinations in continuum $\epem\ra\qqbar$ ($q=u,d,s,c$) events. These
contributions are reduced in part by placing a selection on the
variable $|\cos(\theta_{TB})|$, where $\theta_{TB}$ is the angle
between the thrust axis of the $B$ candidate and the thrust axis of
the rest of the event calculated in the \UfourS\ frame. Candidates
formed in jet-like \qqbar\ events will peak at $|\cos(\theta_{TB})|$
values approaching 1, whereas signal $B$ decays will follow an almost
flat distribution as they are isotropic in this angle. We require
$|\cos(\theta_{TB})|<0.7$ for both $\eta$ channels. The final variable
used in the analysis is a linear Fisher discriminant $\xf$ that
consists of the angles of the $B$ momentum and $B$ thrust axis (in the
\UfourS\ frame) with respect to the beam axis, and the zeroth and
second Legendre moments of the energy flow computed with respect to
the $B$ thrust axis~\cite{PRD}. The reconstruction efficiencies after
selection are presented in Table~\ref{tab:results}.

The analysis uses an extended unbinned maximum-likelihood fit to
extract yields for the modes under study. The input variables to the
fit are $\DE$, $\mes$, $\xf$ and the $a_{0}^{+}$ candidate resonance
mass $m_{\eta\pi}$. The extended likelihood function for the fit is
defined as:
\begin{equation}
 {\cal L} = \frac{e^{-\left(\sum n_j\right)}}{N!} \prod_{i=1}^N \left[\sum_{j=1}^{M}n_j {\cal P}_j\right]\ ,
\end{equation}
where $ {\cal P}_j$ is the normalized PDF for a given fit component
$j$. For each candidate $i$ the PDF is evaluated using the fit
variables of that candidate. The $M$ fit components are the signal and
all background contributions. The total number of candidates is given
by $N$ with the yield associated with each fit component given by
$n_{j}$. The fit for each $\eta$ channel consists of 16 components
modeling signal and continuum candidates separately as well as charged
and neutral charmed $B$ meson decays. There are then 12 components
modeling individual charmless modes which were found to contribute a
background to the signal. The yields for all $B$ background components
are held fixed in the final fit using values calculated from the
latest branching fraction estimates~\cite{PDG2006_hfag}, whereas the
signal and continuum background yields are allowed to vary.

The fit model is constructed in order to extract signal candidates
effectively from a sample where multiple reconstruction hypotheses
exist for each event. The signal MC events have an average candidate
multiplicity of 1.4 for both $\eta$ decay modes. 

In this analysis separate PDFs were used to discriminate between
correctly and incorrectly reconstructed signal candidates in MC. This
was achieved by using MC information to separate the signal MC
candidates into an almost pure sample of correctly reconstructed
candidates and a sample consisting mainly of incorrectly reconstructed
candidates. By iteratively fitting the separate PDFs to each sample in
turn, a consistent set of PDFs for the two cases was obtained. The
component for correctly reconstructed candidates was then taken to
model signal candidates in the final fit to data. The fraction of
events in the MC that were identified as correctly reconstructed by
the fit was approximately 62$\%$ for both $\eta$ channels. The signal
candidate yield resulting from the fit to MC was verified to be
consistent with that expected.

The shapes of the distributions for incorrectly reconstructed signal
were found to be similar to continuum background and thus any such
candidates are assumed to be absorbed into the yield associated with
the continuum PDF. Modeling signal candidates in this way was shown
using ensemble MC studies to provide better sensitivity to signal than
other methods. As a final test, the method was validated using
ensemble MC studies to show that it introduced no bias into the final
fit result.

\begin{table*}[!htbp] 
\begin{small} 
\caption{\small{The results of the fit to the full data set, and other
values required for calculating the branching fraction. All $B$
background yields were held fixed. The upper limit is shown first with
only the statistical error and then with the total error.}}
\vspace{0.1cm} 
\begin{center}
\begin{tabular}{l|c|c} 
\hline
\hline
Required Quantity/Result    & $\eta \to \gamma\gamma$    &  $\eta \to \pip\pim\piz$ \\
\hline\hline
Candidates to fit          & 103054  &  31626 \\
Fixed $B$ Background (candidates) & 1640  & 942   \\
Signal Yield (candidates)      & -8 $\pm$ 19& 13$\pm$13 \\
Continuum Yield (candidates)   & 101400$\pm$300 & 30700$\pm$200 \\
ML Fit Bias (candidates)       &   5.2$\pm$3.0   & $-$2.0$\pm$1.3 \\
\hline
Efficiencies and BFs      & &\\
\hline
Efficiency ($\%$)     &  16.3$\pm$0.1     & 10.2$\pm$0.1 \\
${\cal{B}}(\eta\to X)$ ($\%$)   & 39.4$\pm$0.3  & 22.6$\pm$0.4\\
\vspace{0.01cm}
Branching Fraction  ($\times10^{-6}$) & $-0.6^{+0.8}_{-0.7}$ (stat) $^{+0.4}_{-0.3}$ (syst) & $1.7^{+1.6}_{-1.4}$ (stat) $^{+0.3}_{-0.4}$ (syst) 
\\
\hline
Combined Mode Results                    & \multicolumn{2}{c}{ }\\
\hline
Branching Fraction  ($\times10^{-6}$) & \multicolumn{2}{c}{$0.1^{+0.7}_{-0.7}$ (stat) $^{+0.3}_{-0.3}$ (syst)} \phantom{\Large{T}}\\
Significance & \multicolumn{2}{c}{0.1$\sigma$ (stat + syst)} \\
Upper Limit 90$\%$ C.L. ($\times10^{-6}$) & \multicolumn{2}{c}{$<$ 1.3  (statistical error only)}  \\
Upper Limit 90$\%$ C.L. ($\times10^{-6}$) & \multicolumn{2}{c}{$<$ 1.4 (total error)}  \\
\hline
\hline
\end{tabular}
\label{tab:results}  
\end{center}
\end{small} 
\end{table*}

Any continuum and \BB\ backgrounds that remain after the event
selection criteria have been applied are identified and modeled using
Monte Carlo simulation based on the full physics and detector
models~\cite{geant}. Charmless $B$ decays providing a background to
the signal are identified by analyzing the MC candidates passing
selection from a large mixed sample of Standard Model $B$
decays. Charged and neutral charmed $B$ decays are modeled separately
and individual components are included for each charmless $B$ decay
mode found to contribute. The PDF parameters for each $B$ background
component are obtained from MC samples and held fixed in the final fit
to data. Those for the continuum background shape are left free in the
final fit. The contributions from two charmless backgrounds with the
same final state as signal, those for $B^{+}\to a_{0}(1450)^{+}\piz$
and non-resonant $B^{+}\to\eta\pip\piz$, are estimated using fits to
the relevant regions of the Dalitz plane. Any potential interference
effects were neglected since the fits gave no significant yields for
these modes.

The total PDFs are modeled as products of the PDFs for each of the
four fit variables. The signal shapes in $\DE$, $\mes$, $m_{\eta\pi}$
and $\xf$ are modeled with a Novosibirsk~\cite{NS} function, the sum
of two independent Gaussians, a Breit-Wigner, and an asymmetric
Gaussian, respectively. The signal parameters used for the $a_{0}^{+}$
lineshape are a Breit-Wigner peak value of 983\mevcc with a width of
79\mevcc. These were used in the MC simulation and are consistent with
previous analyses~\cite{teige}, although the width is considered to be
uncertain over a conservative range of 50-100\mevcc in the evaluation
of systematic error. Slowly-varying background distributions in $\xf$
and $m_{\eta\pi}$ are modeled with Chebychev polynomials of the
appropriate order. Such polynomials are also used for $\DE$ in the
charmed $B$ and continuum background cases.  For these components
$\mes$ is modeled with an ARGUS~\cite{argus} threshold function. In
the case of charmless $B$ backgrounds, $\DE$ and $\mes$ are modeled
2-dimensionally using non-parametric PDFs~\cite{keys}, so as to model
correlations between the two variables. Studies of the MC samples for
each mode have shown that these correlations can be as high as 29$\%$.

The results of the analysis are presented in Table~\ref{tab:results}.
The statistical errors on the signal yields are defined using the
change in the central value when the quantity $-2\ln{\cal L}$
increases by one unit from the minimum. The significance is taken as
the square root of the difference between the value of $-2\ln{\cal L}$
for zero signal and the value at the minimum (including additive
systematics).

For the purposes of the branching fraction calculation we assume that
the \UfourS\ decays with an equal rate to both \BpBm\ and
$\BzBzb$~\cite{prodratio}. The fit bias is measured using an ensemble
MC study based on a parameterization taken from the fit to data with
all yield values taken from data. Where a negative yield is found a
value of zero is used for the study. The branching fraction results
from the two $\eta$ decay modes are combined by forming the product of
the likelihood functions, after their maxima have been shifted to
account for fit bias. The functions themselves are defined by
computing the likelihood values for signal yields around the
maximum. Systematic errors are included at the required stages in the
calculation depending on correlations between the two $\eta$ channels.

We find no significant signal in either $\eta$ decay
mode and thus quote upper limits on the branching fraction at the 90\%
confidence level (C.L.), taken to be the branching fraction below
which lies 90\% of the total of the likelihood integral in the
positive branching fraction region.

\begin{figure*}[!ht]
\begin{picture}(100,200)(200,-95)
\put(-1,15){\includegraphics[width=0.245\linewidth]{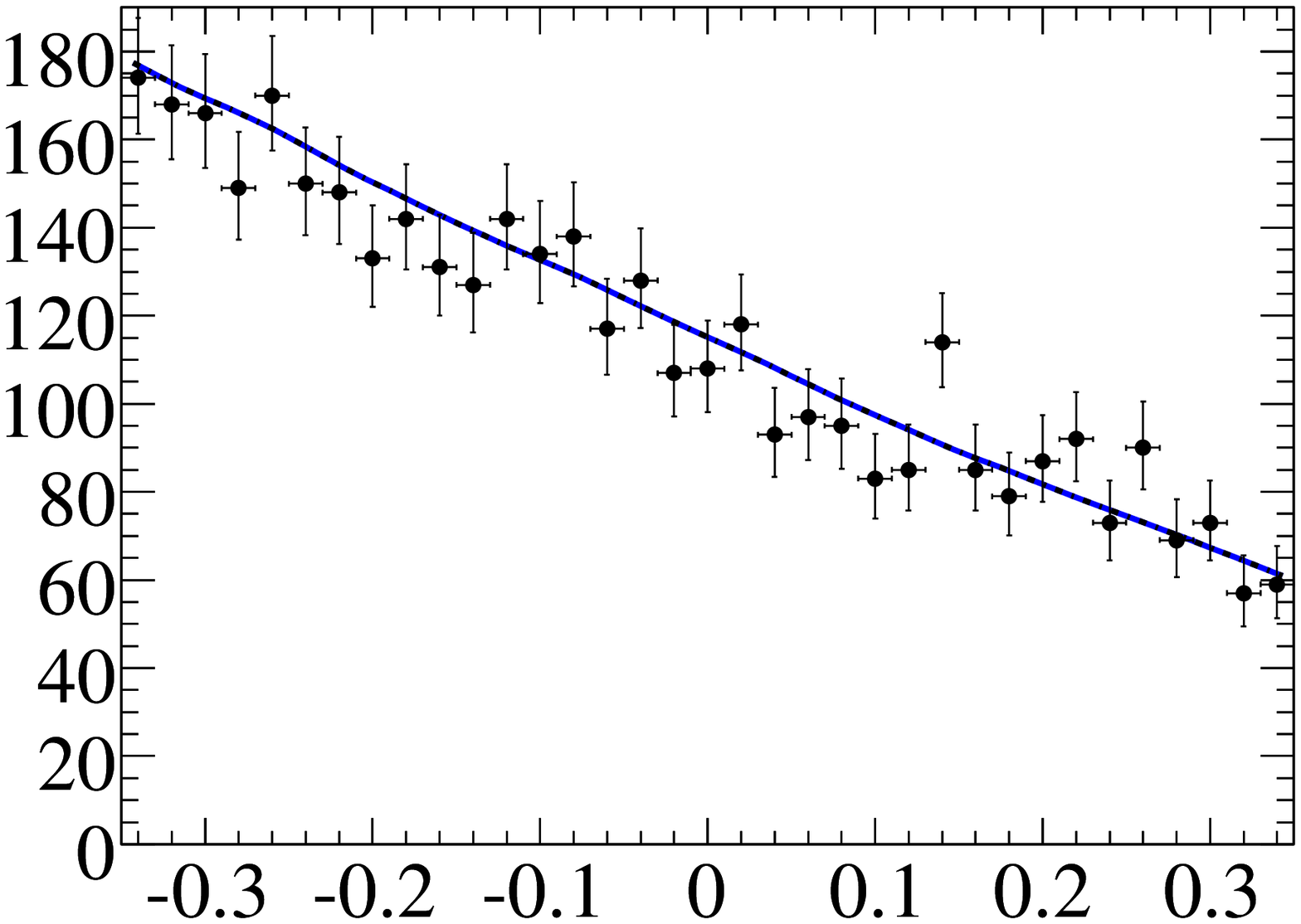}}
\put(-5,34){\rotatebox{90}{{\scriptsize{\bf{Cands./(20{\boldmath{\mev}})}}}}}
\put(65,15){\scriptsize{{\boldmath{$\DeltaE$ ($\gev$)}}}}

\put(126,15){\includegraphics[width=0.245\linewidth]{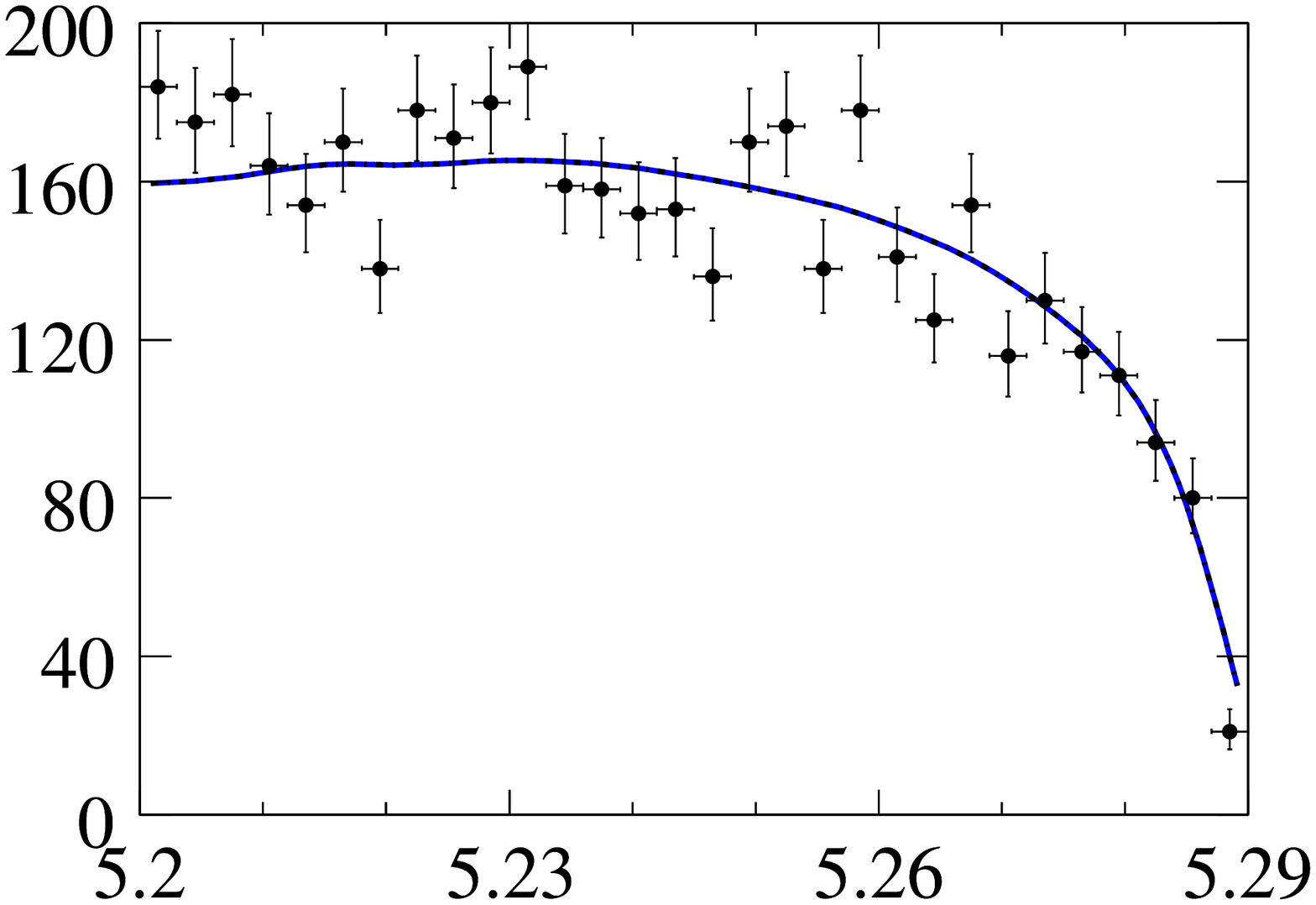}}
\put(120,32){\rotatebox{90}{{\scriptsize{\bf{Cands./(3{\boldmath{\mevcc}})}}}}}
\put(180,15){\scriptsize{{\boldmath{$\mes$ ($\gevcc$)}}}}

\put(-1,-85){\includegraphics[width=0.245\linewidth]{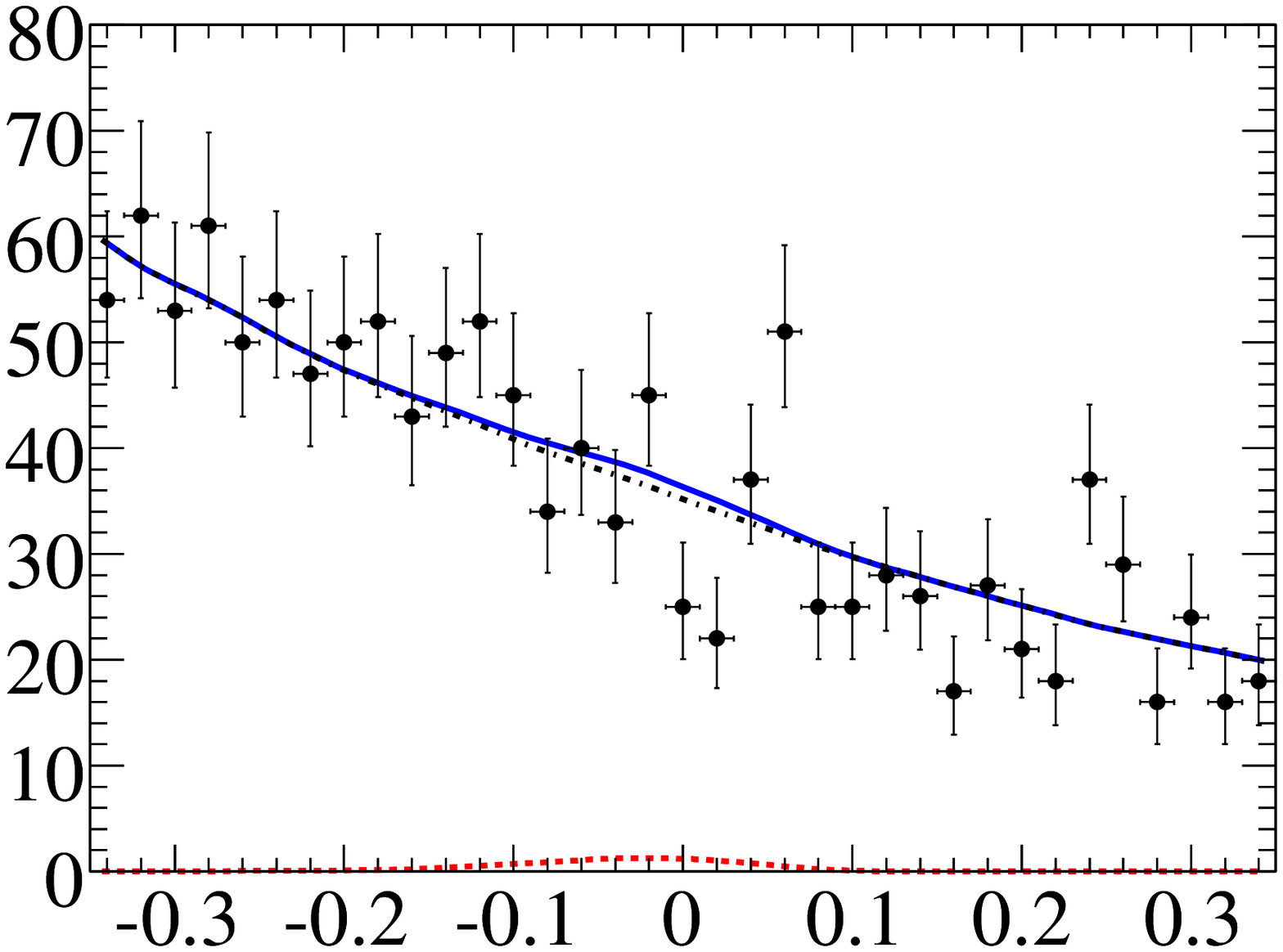}}
\put(-3,-66){\rotatebox{90}{{\scriptsize{\bf{Cands./(20{\boldmath{\mev}})}}}}}
\put(65,-85){\scriptsize{{\boldmath{$\DeltaE$ ($\gev$)}}}}

\put(126,-85){\includegraphics[width=0.245\linewidth]{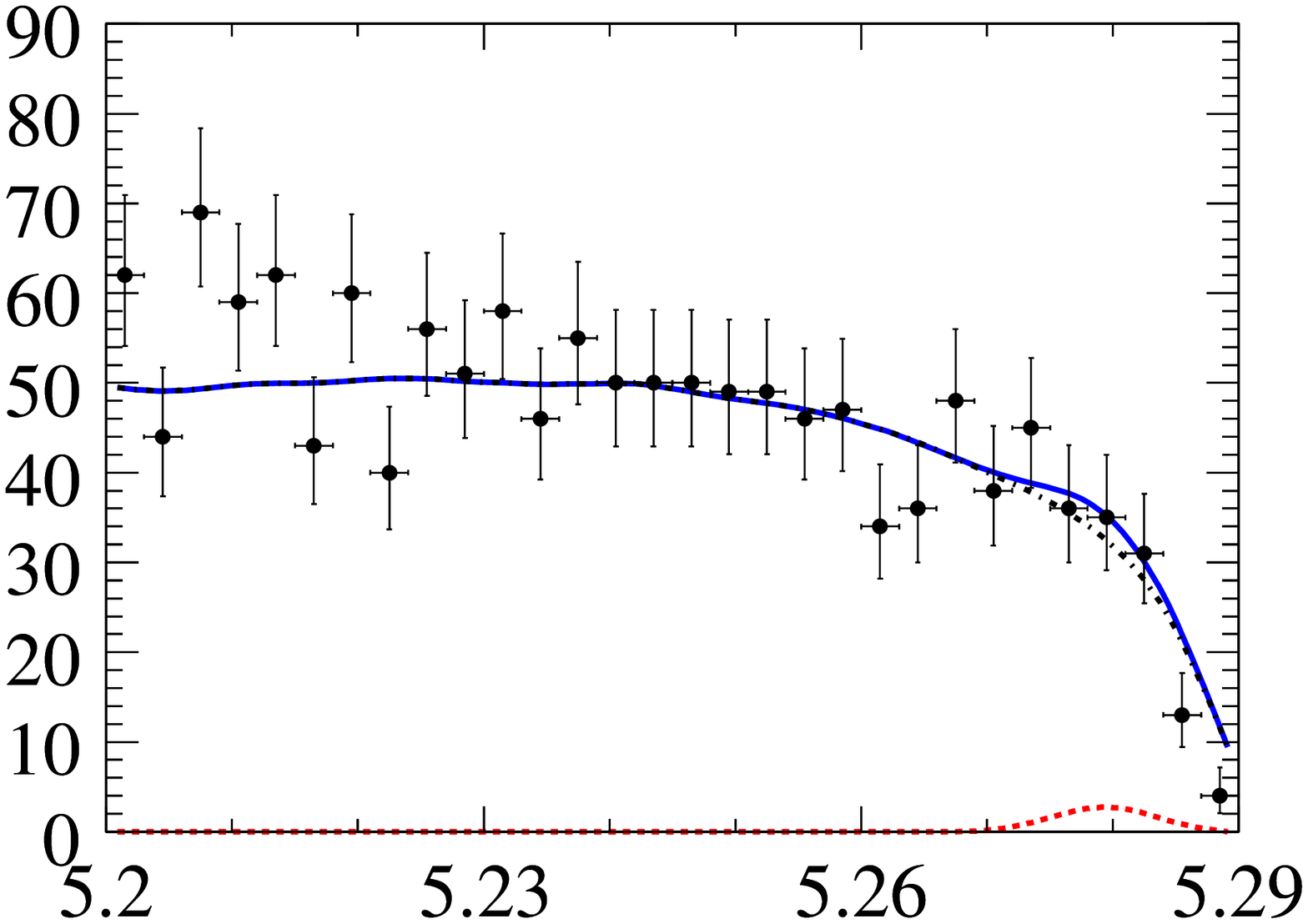}}
\put(122,-70){\rotatebox{90}{{\scriptsize{\bf{Cands./(3{\boldmath{\mevcc}})}}}}}
\put(180,-85){\scriptsize{{\boldmath{$\mes$ ($\gevcc$)}}}}

\put(251,15){\includegraphics[width=0.245\linewidth]{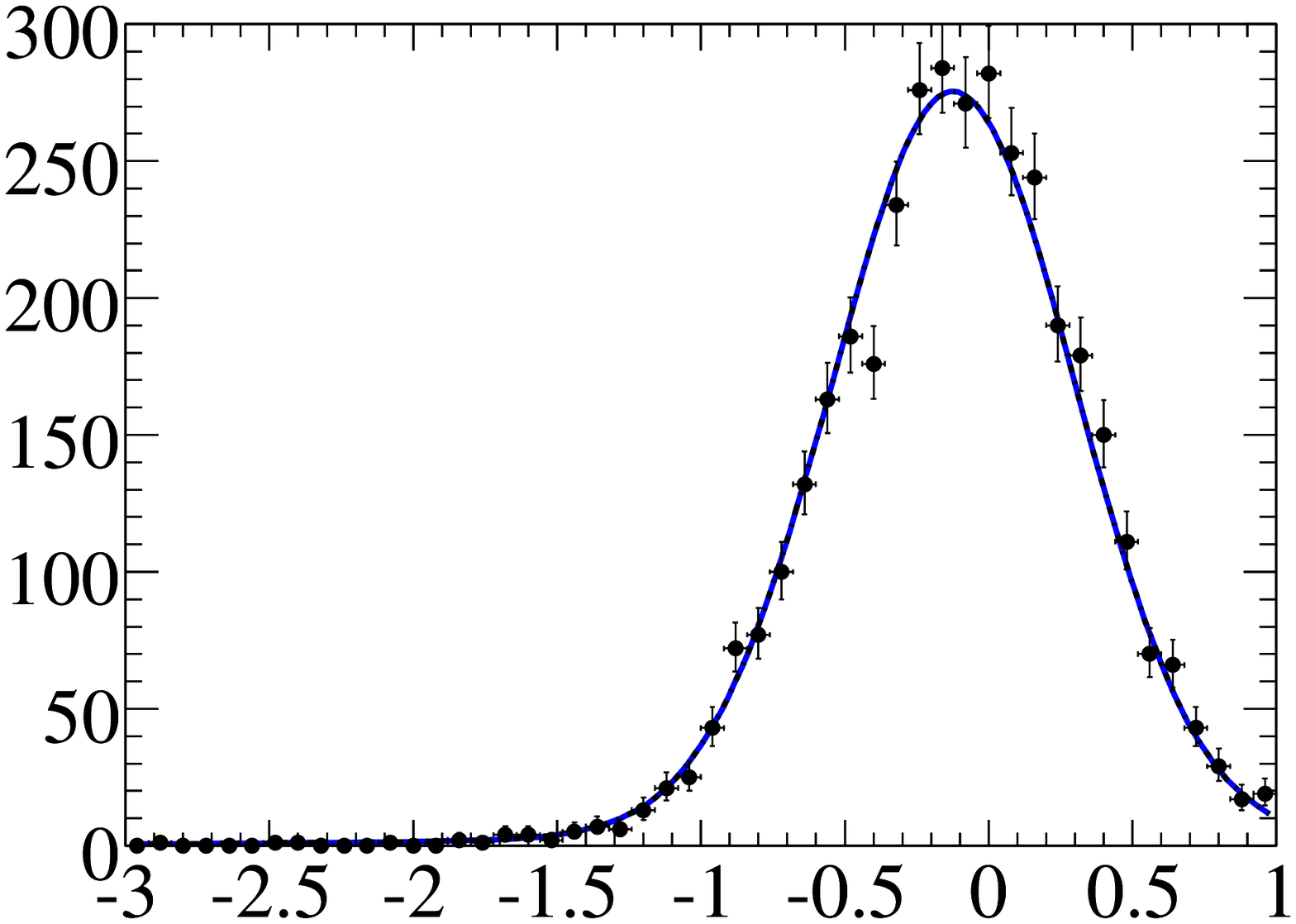}}
\put(248,41){\rotatebox{90}{{\scriptsize{\bf{Cands./(0.08)}}}}}
\put(297,15){\scriptsize{{\boldmath{Fisher discriminant}}}}

\put(379,15){\includegraphics[width=0.245\linewidth]{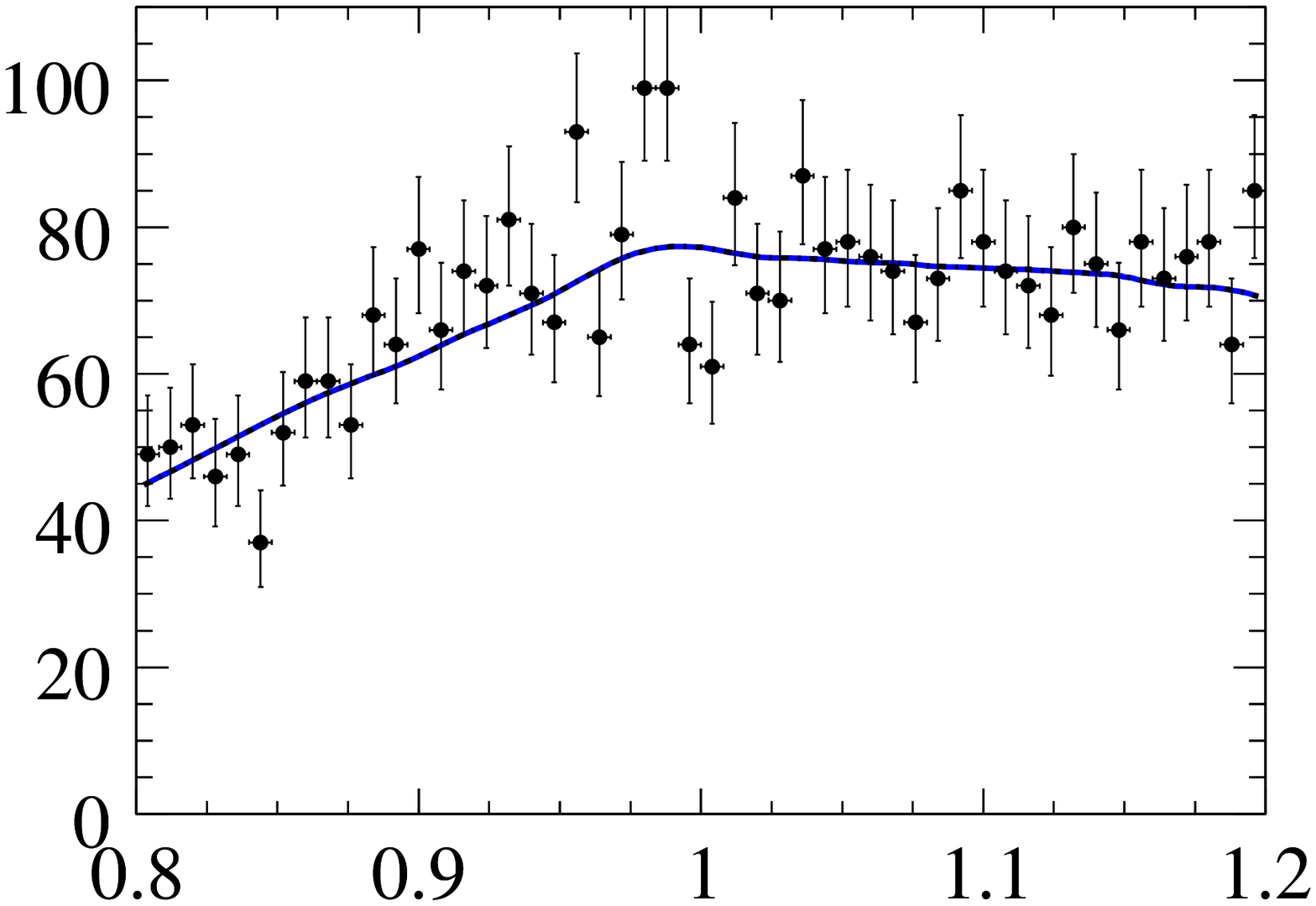}}
\put(372,31){\rotatebox{90}{{\scriptsize{\bf{Cands./(8{\boldmath{\mevcc}})}}}}}
\put(432,15){\scriptsize{{\boldmath{$m_{\eta\pi}$ ($\gevcc$)}}}}

\put(251,-85){\includegraphics[width=0.245\linewidth]{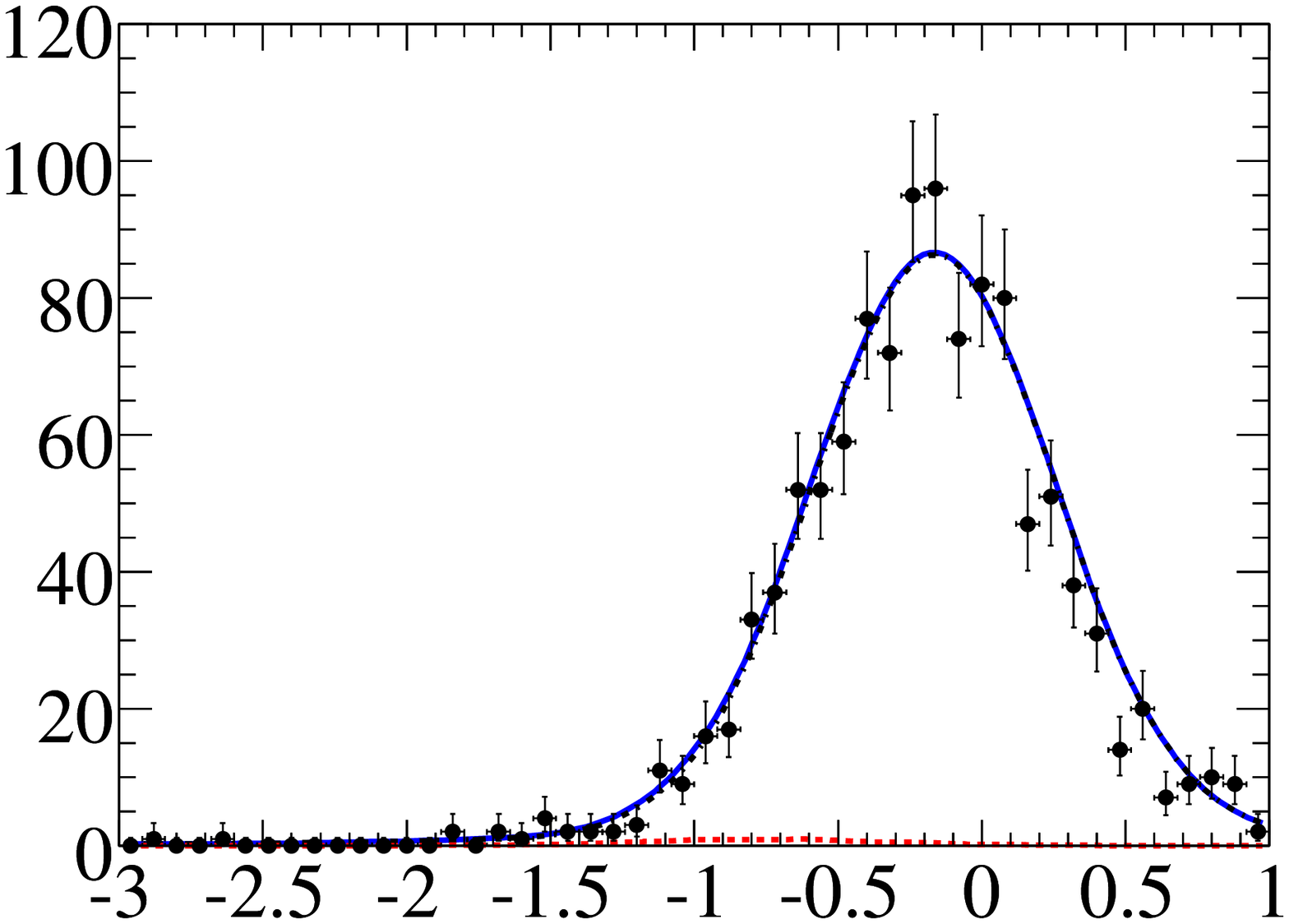}}
\put(248,-61){\rotatebox{90}{{\scriptsize{\bf{Cands./(0.08)}}}}}
\put(297,-85){\scriptsize{{\boldmath{Fisher discriminant}}}}

\put(379,-85){\includegraphics[width=0.245\linewidth]{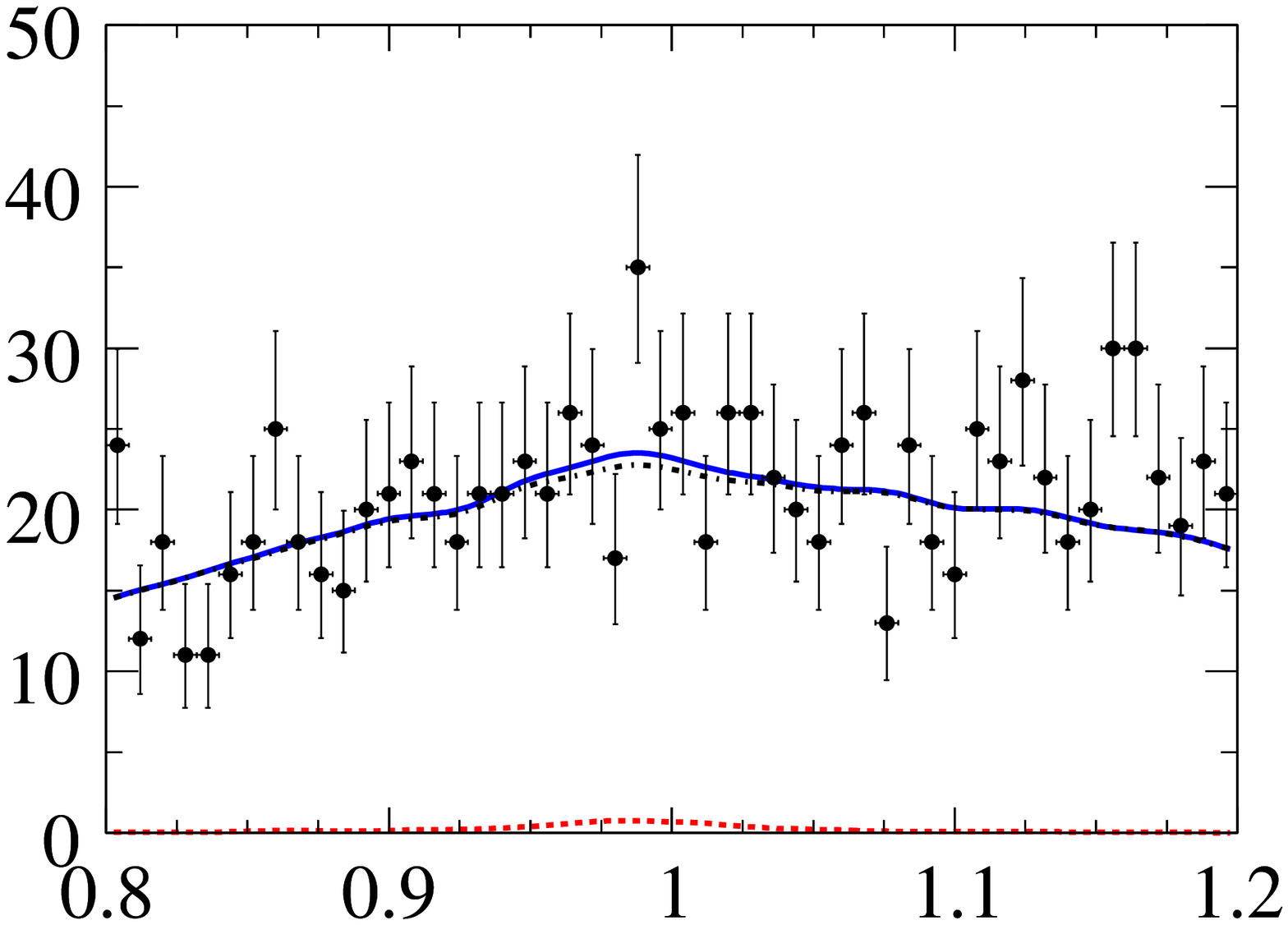}}
\put(374,-69){\rotatebox{90}{{\scriptsize{\bf{Cands./(8{\boldmath{\mevcc}})}}}}}
\put(432,-85){\scriptsize{{\boldmath{$m_{\eta\pi}$ ($\gevcc$)}}}}

\end{picture}
\caption{\label{fig:projplots} Likelihood-ratio-enhanced projections
for the four fit variables (left to right) for the
$\eta\to\gamma\gamma$ (top) and $\eta\to\pip\pim\piz$ (bottom)
cases. Experimental data are represented by points with error bars,
solid blue curves represent the full fit model. For the $\eta_{3\pi}$
case, the combined background component is represented by the black
dash-dotted curve and the signal component by the red dashed
curve. The efficiency of the likelihood ratio selection on the signal
component in the $\eta_{3\pi}$ case is 80.8$\%$.}
\end{figure*}

In Figure~\ref{fig:projplots} we show projections of each of the four
fit variables for both the $\eta\to\gamma\gamma$ and
$\eta\to\pip\pim\piz$ decay modes. To enhance the visibility of a
potential signal, the candidates in these figures have been required
to satisfy the condition that the likelihood ratio
${\cal{L}}_{sig}/[{\cal{L}}_{sig} + \Sigma{\cal{L}}_{bkg}]$ for any
candidate be greater than 0.6. Here ${\cal{L}}_{X}$ is the likelihood
for a given event being described by either the signal or background
model. The likelihoods are calculated for each figure separately,
excluding the variable being plotted. As can be seen there is no
significant signal peak for either mode.

\begin{table}[!ht]
\caption{\small{Estimated systematic errors in the final fit
result. Error sources which are Correlated and Uncorrelated for the
two $\eta$ decay modes are denoted by [C] and [U], respectively.}}
\vspace{0.2cm}
\small
{\centering \begin{tabular}{l|c|c}
\hline\hline
Source of Uncertainty      & $\eta \to \gamma\gamma$     & $\eta \to \pip\pim\piz$ \\
\hline\hline
Additive (Candidates)                     &                   & \\
\hline
Fit Parameters [U] \phantom{\Large{T}}                    & $^{+5.9}_{-4.4}$  & $^{+0.5}_{-1.8}$\\
Charmless Yields [U]                  & $^{+3.6}_{-3.7}$  & $^{+1.2}_{-1.2}$\\
Charm Yields [U]                      & $^{+0.2}_{-0.3}$  & $^{+0.2}_{-0.2}$ \\
Fit Bias [U]                          & $\pm$3.0          & $\pm$1.3\\
\hline
Total Additive (Candidates) \phantom{\Large{T}}               & $^{+7.5}_{-6.5}$  & $^{+1.9}_{-2.6}$\\
\hline\hline
Multiplicative ($\%$)                 &                   &\\
\hline
Neutral Efficiency [C]                & $\pm$6.0 & $\pm$6.0\\
Tracking Efficiency [C]               & $\pm$0.5 & $\pm$1.4\\
$|\cos(\theta_{TB})|$ Selection [C]   & $\pm$3.0 & $\pm$3.0\\
MC Statistics [U]                     & $\pm$0.4 & $\pm$0.3\\
Number of $B\overline{B}$ Events [C]  & $\pm$1.1 & $\pm$1.1\\
Daughter $\eta$ Decay BF [U]          & $\pm$0.7 & $\pm$1.8\\
\hline
Total Multiplicative ($\%$)           & $\pm$6.9 & $\pm$7.2\\
\hline\hline
\vspace{0.03cm}
Total BF Syst Error ($\times10^{-6}$)\phantom{\Large{T}}      & $^{+0.4}_{-0.3}$ & $^{+0.3}_{-0.4}$ \\
\hline\hline
\end{tabular}\par}
\label{tab:systematics}
\end{table}

The largest sources of systematic uncertainty in the analysis arise
from poor knowledge of the $a_{0}^{+}$ lineshape and from the error in
the estimated background contributions. By varying the width of the
$a_{0}^{+}$ Breit-Wigner between 50 and 100\mevcc we predict an
uncertainty of approximately +5 and $-$4 candidates for
$\eta\to\gamma\gamma$ and +0.5 and $-$1 candidate for
$\eta\to\pip\pim\piz$. Varying the charmless yields within their
branching fraction errors (or $\pm$100$\%$ where a limit is used), and
the charmed $B$ yields by $\pm$10$\%$, gives an estimated uncertainty
of $\pm$4 candidates in $\eta\to\gamma\gamma$ and $\pm$1 candidate in
$\eta\to\pip\pim\piz$. The error due to the uncertainty in the fit
bias was calculated as the sum in quadrature of 50$\%$ of the measured
bias and its statistical error, as taken from the ensemble MC study
described above. This value was calculated to be approximately $\pm$3
candidates in the $\eta\to\gamma\gamma$ channel and $\pm$1 candidate
for $\eta\to\pip\pim\piz$.

Further sources of systematic uncertainty, which are multiplicative
rather than additive, affect the efficiency and thus enter into the
branching fraction calculation. Limited signal MC statistics account
for 0.4\% in both $\eta$ decay modes. Auxiliary studies on inclusive
control samples~\cite{PRD}, predict errors of 0.5\% per charged track
and 3\% per reconstructed $\eta$ or \piz decaying to two photons. The
estimate of the number of produced \BB\ events is uncertain by
1.1\%. The uncertainties in $B$ daughter product branching fractions
are taken to be 2\% for $\eta\to\gamma\gamma$ and 3\% for
$\eta\to\pip\pim\piz$~\cite{PDG2006_hfag}. A summary of all systematic
error contributions is presented in Table~\ref{tab:systematics}.

In conclusion, we do not find a significant signal for the mode
$B^{+}\to a_{0}^{+}\piz$. We set an upper limit at 90$\%$ C.L. on the
branching fraction ${\cal{B}}(B^{+}\to
a_{0}^{+}\piz)\times{\cal{B}}(a_{0}^{+}\to\eta\pip)$ of
1.4$\times$10$^{-6}$, suggesting that there is insufficient
sensitivity with the current dataset to probe the predicted
theoretical parameter space, with the largest predicted branching
fraction being 2$\times$10$^{-7}$~\cite{Delepine}. We are therefore
unable to comment on the validity of any of the current models of the
$a_{0}^{+}$.

We are grateful for the excellent luminosity and machine conditions
provided by our \pep2\ colleagues, 
and for the substantial dedicated effort from
the computing organizations that support \babar.
The collaborating institutions wish to thank 
SLAC for its support and kind hospitality. 
This work is supported by
DOE
and NSF (USA),
NSERC (Canada),
IHEP (China),
CEA and
CNRS-IN2P3
(France),
BMBF and DFG
(Germany),
INFN (Italy),
FOM (The Netherlands),
NFR (Norway),
MIST (Russia),
MEC (Spain), and
STFC (United Kingdom). 
Individuals have received support from the
Marie Curie EIF (European Union) and
the A.~P.~Sloan Foundation.


\begin{thebibliography}{99}
\bibitem{CC}
Throughout this paper, charged-conjugate decays are also implied.
\bibitem{NS} The Novosibirsk function is defined as\\ $f(x)=A_{s}
\exp(-0.5(\ln^{2}[1+ \Lambda\tau(x - x_{0})]/\tau^{2} + \tau))$ where
$\Lambda = \sinh(\tau\sqrt{\ln 4})/(\sigma\tau\sqrt{\ln 4})$, the peak
is $x_{0}$, $\tau$ is the tail parameter and $A_{s}$ is a
normalization factor.
\end{thebibliography}
\end{document}